\documentclass[12pt]{iopart}
\pdfoutput=1
\usepackage{cite,amssymb,amsfonts,graphicx,bm,dcolumn,mathrsfs}
\usepackage{color}
\usepackage{ulem}

\begin{document}

\newcommand{\dif}{\rmd}
\newcommand{\ee}{\rme}
\newcommand{\ii}{\rmi}
\newcommand{\kB}{k_\mathrm{B}}
\newcommand{\sub}[1]{\mathrm{#1}}
\newcommand{\vect}[1]{\bi{#1}}

\title{Phase transitions in Thirring's model}
\author{ Alessandro Campa$^1$, Lapo Casetti$^{2,3}$, Ivan Latella$^4$, Agust\'in P\'erez-Madrid$^4$ and Stefano Ruffo$^{2,5}$}
\address{$^1$ Complex Systems and Theoretical Physics Unit, Health and Technology Department, Istituto Superiore di Sanit\`{a},
and INFN Roma 1,
Gruppo Collegato Sanit\`{a}, Viale Regina Elena 299, 00161 Roma, Italy}
\address{$^2$ Dipartimento di Fisica e Astronomia and CSDC, Universit\`a di Firenze,\\and INFN, Sezione di Firenze,
via G.\ Sansone 1, 50019 Sesto Fiorentino (FI), Italy}
\address{$^3$ INAF-Osservatorio Astrofisico di Arcetri, Largo E. Fermi 5, 50125 Firenze, Italy}
\address{$^4$ Departament de F\'{i}sica de la Mat\`eria Condensada, Facultat de F\'{i}sica, Universitat de Barcelona, Mart\'{i} i
Franqu\`{e}s 1, 08028 Barcelona, Spain}
\address{$^5$ SISSA, via Bonomea 265, CNISM and INFN, 34136 Trieste, Italy}
\ead{\mailto{alessandro.campa@iss.infn.it}, \mailto{lapo.casetti@unifi.it}, \mailto{ilatella@ffn.ub.edu},
\mailto{agustiperezmadrid@ub.edu}
and \mailto{ruffo@sissa.it}}

\begin{abstract}
In his pioneering work on negative specific heat, Walter Thirring in\-tro\-duced a model that is solvable in the microcanonical ensemble. Here, we give a complete description of the phase-diagram of this model in both the microcanonical and the canonical ensemble, highlighting the main features of ensemble inequivalence. In both ensembles, we find a line of first-order phase transitions which ends in a critical point. However, neither the line nor the point have the same location in the phase-diagram of the two ensembles. We also show that the microcanonical and canonical critical points can be analytically related to each other using a Landau expansion of entropy and free energy, respectively, in analogy with what has been done in [O. Cohen, D. Mukamel, J. Stat. Mech., P12017 (2012)]. Examples of systems with certain symmetries restricting the Landau expansion have been considered in this reference, while no such restrictions are present in Thirring's model. This leads to a phase diagram that can be seen as a prototype for what happens in systems of particles with kinematic degrees of freedom dominated by long-range interactions.
\end{abstract}


\maketitle

\section{Introduction}

In recent years, the systematic study of systems with long-range interactions has attracted considerable attention, due to remarkable properties that significantly differ from those of short-range interacting systems~\cite{Campa_2014,Campa_2009,Bouchet_2010,Levin_2014}.
Examples of such systems are self-gravitating systems~\cite{Antonov_1962,Lynden-Bell_1968,Lynden-Bell_1999,
Thirring_1970,Padmanabhan_1990,Chavanis_2002,Chavanis_2006,deVega_2002_a,deVega_2002_b}, plasmas~\cite{Nicholson_1992,Kiessling_2003},
two-dimensional and geophysical fluids~\cite{Onsager_1949,Miller_1990,Robert_1991,Chavanis_2002_b,Venaille_2012} and
spin systems~\cite{Mori_2010,Mori_2011}. The long-range character of the interactions confers a striking property to these
systems: they are intrinsically non additive. Non additivity, however, does not hinder neither a statistical mechanical
formulation~\cite{Campa_2014} nor a proper thermodynamic description~\cite{Latella_2015}.
Because of non additivity, equilibrium configurations may present negative specific heat in the microcanonical
ensemble~\cite{Lynden-Bell_1968,Thirring_1970}, ensemble inequivalence~\cite{Thirring_1970,Ellis_2000,Barre_2001,Bouchet_2005}
and the violation of the usual Gibbs-Duhem equation~\cite{Latella_2013,Latella_2015}.
A feature which is of direct relevance for this paper is that non additivity, which is responsible for changes in the concavity of the thermodynamic potentials, directly leads to ensemble inequivalence. This latter is in turn manifest through the properties of the phase-diagrams, which are not the same in different ensembles.  

A seminal work on negative specific heat was written by Walter Thirring~\cite{Thirring_1970}. In that paper he
introduced a simple model that reproduces some of the properties of self-gravitating systems. He showed that
the model exhibits negative specific heat and temperature jumps in the microcanonical ensemble and that they are
both absent, and replaced by a first-order phase transition, in the canonical ensemble. In the last decade, 
ensemble inequivalence in non additive systems has become an established fact~\cite{Campa_2014}, and several
different models have been shown to display such a feature. 

However, quite surprisingly, a detailed study of the full phase diagram in both the microcanonical and 
the canonical ensemble of Thirring's model has not yet been performed. Moreover, the analysis of ensemble
inequivalence has been restricted in general to models which are endowed with specific symmetries of the order 
parameter. We are here thinking, for instance, to models of magnetic systems which, in absence of an external field, are invariant under a sign change $m \to -m$ of magnetization $m$. These models have typically a phase diagram 
with a line of second order phase transitions which ends at a tricritical point. This latter has a different location in different ensembles~\cite{Barre_2001}. Thirring's model does not possess this symmetry and, as we will show, the line
of first-order phase transitions terminates at a critical point, as it happens for the gas-liquid phase
transitions in fluids. At variance with what is found for models with symmetries, ensemble inequivalence
manifests itself in Thirring's model by a different location of the critical point. In addition, the mean-field character of Thirring's model allows us to employ a Landau expansion~\cite{Landau_1980} of thermodynamic potentials and to determine analytically the location of the critical point in both the microcanonical
and canonical ensembles.

Using a Landau expansion for the thermodynamic potentials, phase diagrams and ensemble inequivalence in systems with two types of symmetry have been considered in~\cite{Cohen_2012}. There, these symmetries are specified by $f(m,q)=f(-m,q)$ and $f(m,q)=f(-m,-q)$, where $m$ is the order parameter and $f$ is the thermodynamic potential corresponding to the ``lower'' ensemble in which the thermodynamic variable $q$ is fixed, while this variable can fluctuate in the ``higher'' ensemble. Furthermore,
the ABC model~\cite{Lederhendler_2010,Lederhendler_2010_2}, a one-dimensional driven exclusion model, and the anisotropic XY model for a system with infinite-range interacting spins have been discussed in~\cite{Cohen_2012} as concrete examples. These models consist of lattice sites with internal degrees of freedom; on the contrary, Thirring's model is a simplified version of a self-gravitating system, and as such it describes particles with kinematic degrees of freedom. In addition and in contrast to the previous examples, Thirring's model has no symmetries restricting the Landau expansion, leading to a phase diagram that can be seen as a prototype for what happens in systems of particles with kinematic degrees of freedom dominated by long-range interactions.

\section{Thirring's model}

Thirring's model is a minimal model that describes a confined system with regularized attractive interactions that mimic those of a
self-gravitating gas~\cite{Casetti_a,Casetti_b}. In this model, $N$ particles of mass $m$ are enclosed in a volume $V$ with a
Hamiltonian given by
\begin{equation}
\mathcal{H}=\sum_{i=1}^N\frac{|\vect{p}_i|^2}{2m}+\sum_{i> j}^N\phi(\vect{q}_i,\vect{q}_j),
\end{equation}
where $\vect{p}_i$ is the momentum of the $i$-th particle, and the interactions are defined by the nonlocal potential~\cite{Thirring_1970}
\begin{equation}
\phi(\vect{q}_i,\vect{q}_j)=-2\nu\theta_{V_0}(\vect{q}_i) \theta_{V_0}(\vect{q}_j). 
\label{potential}
\end{equation}
Here $\nu>0$ is a constant, and $\theta_{V_0}(\vect{q}_i)=1$ if $\vect{q}_i\in V_0$ and vanishes otherwise, where $\vect{q}_i$
is the position of the $i$-th particle and $V_0< V$ is the core volume. Particles outside $V_0$ are free, so that the total potential energy
in the large $N$ limit is given by
\begin{equation}
\sum_{i> j}^N\phi(\vect{q}_i,\vect{q}_j)=-\nu N_0^2,
\label{potenergy}
\end{equation}
where $N_0$ is the number of particles in $V_0$ for a given configuration. Notice that, as a consequence of the interaction
potential~(\ref{potential}), the system is nonadditive~\cite{Latella_2015} and exhibits the rich phenomenology common to long-range
interacting systems. In particular, the microcanonical and canonical ensembles are not equivalent, as will be shown below.

Let us consider the thermodynamics of the system when it is isolated. The density of states in phase space can be written as a sum
over all possible values of the number of particles in the core~\cite{Thirring_1970}
\begin{equation}
\omega(E,V,N)= \sum_{N_0}\ee^{\hat{S}(E,V,N,N_0)},
\label{density_of_states}
\end{equation}
in such a way that the maximization of $\hat{S}(E,V,N,N_0)$ leads to the microcanonical entropy in the large $N$ limit,
$S(E,V,N)=\sup_{N_0}\hat{S}(E,V,N,N_0)$. Here and below we use units in which $\kB=1$. Furthermore, introducing the fraction of
free particles $n_\sub{g}$ (fraction of particles outside $V_0$), the reduced energy $\varepsilon$, and the reduced volume $\eta$,
given by
\begin{equation}
n_\sub{g}=1-\frac{N_0}{N},\qquad\varepsilon=\frac{E}{\nu N^2}+1,\qquad\eta=\ln\left(\frac{V-V_0}{V_0}\right),
\end{equation}
the function $\hat{S}$ in (\ref{density_of_states}) can be written as $\hat{S}\equiv N\hat{s}$ with~\cite{Thirring_1970,Casetti_b}
\begin{equation}
\fl\hat{s}(n_\sub{g},\varepsilon,\eta)=\frac{3}{2}\ln\left[\varepsilon-2n_\sub{g}+n_\sub{g}^2\right] 
-(1-n_\sub{g})\ln(1-n_\sub{g})-n_\sub{g}\ln n_\sub{g}+n_\sub{g}\eta, 
\label{entropy}
\end{equation}
where in (\ref{entropy}) we have neglected constant terms. The microcanonical entropy per particle $s=S/N$ is thus given by
\begin{equation}
s(\varepsilon,\eta)= \hat{s}(\bar{n}_\sub{g},\varepsilon,\eta)=\sup_{n_\sub{g}} \hat{s}(n_\sub{g},\varepsilon,\eta),
\end{equation}
where $\bar{n}_\sub{g}=\bar{n}_\sub{g}(\varepsilon,\eta)$ is the value of $n_\sub{g}$ that maximizes (\ref{entropy}).
The energy $E$, being the sum of the potential energy (\ref{potenergy}) and of the kinetic energy $\mathcal{K}$, is bounded from below by
$-\nu N^2$, therefore for the reduced energy we have $\varepsilon \ge 0$. Furthermore, for a given reduced energy in the range
$0 \le \varepsilon < 1$, the fraction of free particles $n_\sub{g}$ is bounded from above by $1-\sqrt{1-\varepsilon}$, due to the fact that $\mathcal{K}\ge 0$. On the other hand, for $\varepsilon \ge 1$ the fraction $n_\sub{g}$ can take any value in the range
$0\le n_\sub{g} \le 1$. In turn, this guarantees that the argument of the logarithm in equation (\ref{entropy}) is never negative.
The reduced temperature $\tau=T/(\nu N)$, where $T$ is the temperature, takes the form 
\begin{equation}
\frac{1}{\tau(\varepsilon,\eta)}=\left.\frac{\partial }{\partial \varepsilon}\hat{s}(n_\sub{g},\varepsilon,\eta)\right|_{n_\sub{g}=
\bar{n}_\sub{g}}=\frac{3}{2}\left(\varepsilon-2\bar{n}_\sub{g}+\bar{n}_\sub{g}^2\right)^{-1},
\label{inverse_temperature_micro}
\end{equation}
which is guaranteed to be positive from the same observation made above.

In the canonical ensemble, the system is assumed to be in contact with a thermostat, in such a way that the reduced temperature
$\tau$ is fixed and the energy fluctuates. The reduced canonical free energy is $\varphi=F/(NT)$, $F$ being the canonical free
energy. It can be obtained from the microcanonical entropy by computing its Legendre-Fenchel transform~\cite{Campa_2014}, namely,
\begin{equation}
\varphi(\tau,\eta)=\inf_{\varepsilon} \left[\frac{\varepsilon}{\tau}- s(\varepsilon,\eta)\right].
\label{Legendre-Fenchel}
\end{equation}
The reduced free energy can also be written as
\begin{equation}
\varphi(\tau,\eta)=\hat{\varphi}(\bar{n}_\sub{g},\tau,\eta)=\inf_{n_\sub{g}}\hat{\varphi}(n_\sub{g},\tau,\eta),
\end{equation}
where 
\begin{equation}
\hat{\varphi}(n_\sub{g},\tau,\eta)=\inf_\varepsilon\left[\frac{\varepsilon}{\tau}- \hat{s}(n_\sub{g},\varepsilon,\eta)\right],
\label{free_energy2}
\end{equation}
and now the fraction of free particles that minimizes the free energy is a function of the temperature,
$\bar{n}_\sub{g}=\bar{n}_\sub{g}(\tau,\eta)$. In this case, using (\ref{entropy}), the expression (\ref{free_energy2}) can
be computed to give
\begin{equation}
\fl\hat{\varphi}(n_\sub{g},\tau,\eta)=-\frac{3}{2}\ln\left(\frac{3\tau}{2}\right)+\frac{2n_\sub{g}-n_\sub{g}^2}{\tau} 
+(1-n_\sub{g})\ln(1-n_\sub{g}) +n_\sub{g}\ln n_\sub{g}-n_\sub{g}\eta+\frac{3}{2}.
\label{free_energy}
\end{equation}
Obviously, the constant terms neglected in the entropy (\ref{entropy}) are not included. The mean value
$\bar{\varepsilon}$ of the reduced energy in the canonical ensemble is given by
\begin{equation}
\bar{\varepsilon}(\tau,\eta)=\left.-\tau^2\frac{\partial }{\partial \tau}\hat{\varphi}(n_\sub{g},\tau,\eta)\right|_{n_\sub{g}=
\bar{n}_\sub{g}}=\frac{3\tau}{2}+2\bar{n}_\sub{g}-\bar{n}_\sub{g}^2. 
\label{mean_energy}
\end{equation}

An interesting feature of the system is that it undergoes first-order phase transitions in both the microcanonical and canonical
ensembles. Using the Landau theory of phase transitions, below we study the critical points in the two ensembles and show explicitly
that they differ from each other.

\section{Landau theory: Microcanonical ensemble}

Let us introduce the deviation $m=n_\sub{g}-\bar{n}_\sub{g}$ of the fraction of free particles $n_\sub{g}$ with respect to a
certain reference value $\bar{n}_\sub{g}$. This reference value will be the one maximizing equation (\ref{entropy}), i.e., the
equilibrium value.
Thus, we perform a Landau expansion of the entropy (\ref{entropy}) in powers of $m$ around $\bar{n}_\sub{g}$,
\begin{eqnarray}
\fl \hat{s}(m,\varepsilon,\eta)=&a_s(\bar{n}_\sub{g},\varepsilon,\eta)+b_s(\bar{n}_\sub{g},
\varepsilon,\eta)m+c_s(\bar{n}_\sub{g},\varepsilon,\eta)m^2
+d_s(\bar{n}_\sub{g},\varepsilon,\eta)m^3\nonumber\\
&+e_s(\bar{n}_\sub{g},\varepsilon,\eta)m^4+\mathcal{O}(m^5),
\label{Landau_entropy}
\end{eqnarray}
where the coefficients are given by
\begin{eqnarray}
\fl a_s(\bar{n}_\sub{g},\varepsilon,\eta)&=&\frac{3}{2} \ln \left(\varepsilon-2\bar{n}_\sub{g}+ \bar{n}_\sub{g}^2\right)-
(1- \bar{n}_\sub{g}) \ln (1- \bar{n}_\sub{g})- \bar{n}_\sub{g} \ln  \bar{n}_\sub{g}+\bar{n}_\sub{g}  \eta,\label{a_s}\\
\fl b_s(\bar{n}_\sub{g},\varepsilon,\eta)&=&\ln\left(\frac{1-\bar{n}_\sub{g}}{\bar{n}_\sub{g}}\right)
-\frac{3 (1- \bar{n}_\sub{g})}{\varepsilon-2\bar{n}_\sub{g}+ \bar{n}_\sub{g}^2}+ \eta\label{b_s},\\
\fl c_s(\bar{n}_\sub{g},\varepsilon,\eta)&=&
\frac{P_\varepsilon(\bar{n}_\sub{g})}{2\left(1-\bar{n}_\sub{g}\right)\bar{n}_\sub{g}\left(\varepsilon-2\bar{n}_\sub{g}+
 \bar{n}_\sub{g}^2\right)^2}\label{c_s},\\
\fl d_s(\bar{n}_\sub{g},\varepsilon,\eta)&=&\frac{1-2  \bar{n}_\sub{g}}{6 (1- \bar{n}_\sub{g})^2  \bar{n}_\sub{g}^2}
+\frac{3 (1- \bar{n}_\sub{g})}{\left(\varepsilon-2\bar{n}_\sub{g}+ \bar{n}_\sub{g}^2\right)^2}
-\frac{4  (1-\bar{n}_\sub{g})^3}{\left(\varepsilon-2\bar{n}_\sub{g}+ \bar{n}_\sub{g}^2\right)^3},\label{d_s}\\
\fl e_s(\bar{n}_\sub{g},\varepsilon,\eta)&=&-\frac{\bar{n}_\sub{g}^3+(1- \bar{n}_\sub{g})^3}{12(1- \bar{n}_\sub{g})^3\bar{n}_\sub{g}^3}
-\frac{3}{4\left(\varepsilon-2\bar{n}_\sub{g}+ \bar{n}_\sub{g}^2\right)^2}
-\frac{6 (1-\varepsilon) (1-\bar{n}_\sub{g})^2}{\left(\varepsilon-2\bar{n}_\sub{g}+ \bar{n}_\sub{g}^2\right)^4},\label{e_s}
\end{eqnarray}
with
\begin{equation}
P_\varepsilon(\bar{n}_\sub{g})\equiv2 \bar{n}_\sub{g}^4-5\bar{n}_\sub{g}^3+(8-5\varepsilon)\bar{n}_\sub{g}^2
+(7\varepsilon-6)\bar{n}_\sub{g}-\varepsilon^2.
\label{polynomial}
\end{equation}

We note that the equilibrium states require the conditions $b_s(\bar{n}_\sub{g},\varepsilon,\eta)=0$, defining
$\bar{n}_\sub{g}=\bar{n}_\sub{g}(\varepsilon,\eta)$, and $c_s(\bar{n}_\sub{g},\varepsilon,\eta) \le 0$.
It is not difficult to see that for $0\le \varepsilon \le 1$ these conditions are satisfied by only one value of
$\bar{n}_\sub{g}$; therefore a phase transition can occur only for $\varepsilon >1$.

\begin{figure}
\centering
\includegraphics[scale=1]{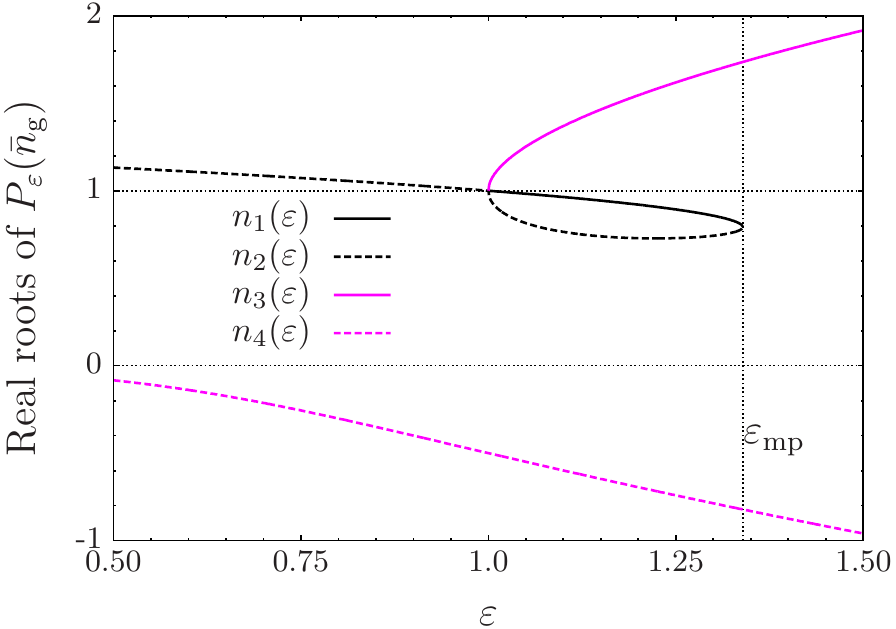} 
\caption{Real roots of $P_\varepsilon(\bar{n}_\sub{g})$ as a function of the reduced energy $\varepsilon$. Since, by definition,
$0\leq\bar{n}_\sub{g}\leq1$,
the roots $n_3(\varepsilon)$ and $n_4(\varepsilon)$ are not to be considered. The roots $n_1(\varepsilon)$ and
$n_2(\varepsilon)$ become equal at $\varepsilon_\sub{mp}$, the energy at the microcanonical critical point.}
\label{graph_roots}
\end{figure}

\subsection{Microcanonical critical point}

The microcanonical phase diagram in the $(\varepsilon,\eta)$ plane exhibits a line of first-order phase transition
that ends at a critical point specified by the reduced energy and volume $\varepsilon_\sub{mp}$ and $\eta_\sub{mp}$,
respectively, corresponding to a fraction of free particles $\bar{n}_\sub{g}=n_\sub{mp}$. Such critical values can be obtained
by solving the system of equations
\begin{equation}
\eqalign{
b_s(n_\sub{mp},\varepsilon_\sub{mp},\eta_\sub{mp})=0,\\
c_s(n_\sub{mp},\varepsilon_\sub{mp},\eta_\sub{mp})=0,\\
d_s(n_\sub{mp},\varepsilon_\sub{mp},\eta_\sub{mp})=0.
}
\label{micro_system_equations}
\end{equation}

In order to find the critical point, consider the quartic polynomial $P_\varepsilon(\bar{n}_\sub{g})$ given by (\ref{polynomial});
when $P_\varepsilon(\bar{n}_\sub{g})$ vanishes, also the coefficient $c_s$ vanishes. Let us denote the roots of
$P_\varepsilon(\bar{n}_\sub{g})$ by $n_i(\varepsilon)$, $i=1,\ldots,4$. Two of these roots, say, $n_3(\varepsilon)$ and
$n_4(\varepsilon)$, lie outside the interval $[0,1)$ when they are real: since the fraction $\bar{n}_\sub{g}$ is bounded,
$0\leq\bar{n}_\sub{g}\leq1$, these roots are not to be considered. The other two roots, $n_1(\varepsilon)$ and $n_2(\varepsilon)$,
can be real or complex, depending on the value of $\varepsilon$, and are given by
\begin{eqnarray}
&n_1(\varepsilon)=\frac{\sqrt{3}}{24}
\left\{\frac{15}{\sqrt{3}}+ z_2(\varepsilon )- \left[2z_1(\varepsilon )-18 \sqrt{3}\frac{ (8 \varepsilon +1)}
{z_2(\varepsilon )}\right]^{1/2}\right\}, \label{n_1}\\
&n_2(\varepsilon)=\frac{\sqrt{3}}{24}
\left\{\frac{15}{\sqrt{3}}- z_2(\varepsilon )+ \left[2z_1(\varepsilon )+18 \sqrt{3}\frac{ (8 \varepsilon +1)}
{z_2(\varepsilon )}\right]^{1/2}\right\}, \label{n_2}
\end{eqnarray}
where
\begin{eqnarray}
z_1(\varepsilon)&=16 (5 \varepsilon -8)-\frac{4 (\varepsilon -1) (\varepsilon +26)}{z_3(\varepsilon )}-4 z_3(\varepsilon )+75,\\
z_2(\varepsilon)&=\left[80 \varepsilon +8 z_3(\varepsilon )+\frac{8 (\varepsilon -1) 
(\varepsilon +26)}{z_3(\varepsilon )}-53\right]^{1/2},\\
z_3(\varepsilon)&=\left[\frac{3 \sqrt{3}}{2}\sqrt{-\Delta_P(\varepsilon )}+  (1374-485 \varepsilon )
 \varepsilon^2 -1293\varepsilon+404\right]^{1/3},
\end{eqnarray}
and
\begin{equation}
\Delta_P(\varepsilon )=-36 (\varepsilon -1)^3 \left(  968 \varepsilon^3 -2581\varepsilon^2+2276\varepsilon -744\right)
\label{discriminant}
\end{equation}
is the discriminant of $P_\varepsilon(\bar{n}_\sub{g})$. When $n_1(\varepsilon)$ and $n_2(\varepsilon)$ are real, they lie
in the interval $(0,1]$ for a certain range of energies $\varepsilon$. To visualize this situation,
we plot in figure \ref{graph_roots} the real roots of $P_\varepsilon(\bar{n}_\sub{g})$ as function of $\varepsilon$.
In addition, these roots are real and different when the discriminant is positive, are real and degenerate when
$\Delta_P(\varepsilon )$ vanishes, and become complex when $\Delta_P(\varepsilon )$ is negative.
Thus, the solution of the system (\ref{micro_system_equations}) is characterized by the
condition $\Delta_P(\varepsilon_\sub{mp})=0$, in such a way that $n_\sub{mp}=n_1(\varepsilon_\sub{mp})=n_2(\varepsilon_\sub{mp})$. 
This can be seen by noting that $c_s$ is continuous for $\bar{n}_\sub{g}$ between $n_1$ and $n_2$, and that
\begin{equation}
d_s(\bar{n}_\sub{g},\varepsilon,\eta)=\frac{1}{3}\frac{\partial }{\partial\bar{n}_\sub{g}}c_s(\bar{n}_\sub{g},\varepsilon,\eta), 
\end{equation}
so that the value of $\bar{n}_\sub{g}$ that cancels out $d_s$ must lie between $n_1$ and $n_2$. Therefore, if the fraction
$\bar{n}_\sub{g}=n_\sub{mp}$ cancels out both $c_s$ and $d_s$, we have $n_\sub{mp}=n_1=n_2$, which is precisely what happens
when the discriminant vanishes, $\Delta_P(\varepsilon_\sub{mp})=0$. Furthermore, in such a case, from
$b_s(n_\sub{mp},\varepsilon_\sub{mp},\eta_\sub{mp})=0$, the critical reduced volume can be unequivocally
determined as 
\begin{equation}
\eta_\sub{mp}= \frac{3 \left(1-n_\sub{mp}\right)}{\varepsilon_\sub{mp}-2 n_\sub{mp}+n_\sub{mp}^2}-
\ln\left(\frac{1- n_\sub{mp}}{ n_\sub{mp}}\right).
\label{eta_mp}
\end{equation}

The discriminant $\Delta_P(\varepsilon )$, equation (\ref{discriminant}), is a polynomial of degree six in $\varepsilon$. It
has four reals roots and two complex roots: three of these real roots are found at $\varepsilon_0=1$, and the remaining real
root is given by
\begin{equation}
\varepsilon_\sub{mp}=\frac{\left(k_1+k_2\right)^{1/3}+\left(k_1-k_2\right)^{1/3}+2581}{2904}\simeq 1.339,
\label{critvol}
\end{equation}
with the numerical coefficients $k_1=1016263261$ and $k_2=37792656 \sqrt{723}$. We note that $\varepsilon_0$ is not the
critical energy at the critical point, since one has $n_1(\varepsilon_0)=n_2(\varepsilon_0)=1$ and, hence, this corresponds,
from equation (\ref{eta_mp}), to a state with $\eta\to\infty$. The critical fraction can
be obtained by evaluating equations (\ref{n_1}) or (\ref{n_2}) at the critical energy $\varepsilon_\sub{mp}$, yielding
$n_\sub{mp}\simeq0.7929$. Finally, from (\ref{eta_mp}), the critical reduced volume is given by $\eta_\sub{mp}\simeq2.969$.

\begin{figure}
\centering
\includegraphics[scale=1]{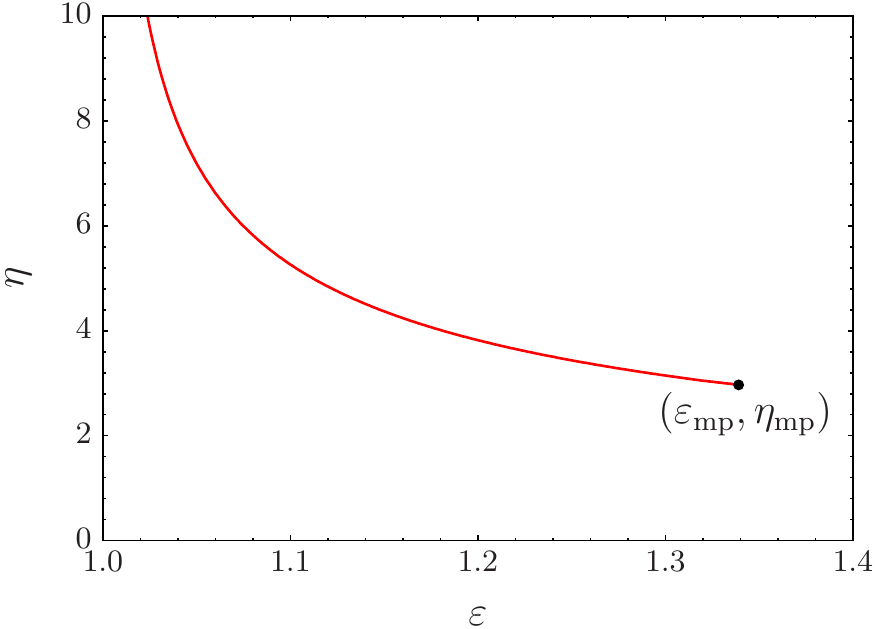} 
\caption{Phase diagram in the $(\varepsilon,\eta)$ plane, showing the transition line in the microcanonical ensemble. The plot shows the
curve $\eta(\varepsilon)$, which represents the points defined by the reduced energy $\varepsilon$ and reduced volume
$\eta$ at which first-order phase transitions take place in the microcanonical ensemble. The transition
line terminates at a critical point represented by the point $(\varepsilon_\sub{mp},\eta_\sub{mp})$. The values of the critical
parameters are $\varepsilon_\sub{mp}\simeq 1.339$ and $\eta_\sub{mp}\simeq2.969$.}
\label{micro_phasediagram_e}
\end{figure}

In addition, we note that phase transitions can occur only for $\varepsilon$ such that
$\varepsilon_0<\varepsilon<\varepsilon_\sub{mp}$, since for $\varepsilon<1$, as noted before, the condition
$b_s(\bar{n}_\sub{g},\varepsilon,\eta)=0$ defines only one state of equilibrium.
In figure \ref{micro_phasediagram_e} we show the microcanonical phase diagram in the $(\varepsilon,\eta)$ plane, with the line
of first order transition points terminating at the critical point. 
The features of the microcanonical and canonical phase diagrams are commented later.

\section{Landau theory: Canonical ensemble}

We are interested in showing how the phase diagram in the canonical ensemble differs from the diagram obtained in the microcanonical
ensemble. Following~{\cite{Cohen_2012}, we introduce the deviation $q=\varepsilon-\bar{\varepsilon}$ of the energy with respect
to the mean value $\bar{\varepsilon}$ and perform an expansion in powers of $q$ of the entropy in such a way that
\begin{eqnarray}
\fl\frac{\varepsilon}{\tau}-\hat{s}(m,\varepsilon,\eta)=&\frac{\bar{\varepsilon}+q}{\tau}-a_0-a_1 q-a_2q^2-a_3q^3-
\left(b_0+b_1q+b_2q^2\right)m\nonumber\\
&-\left(c_0+c_1q\right)m^2-d_0m^3+\mathcal{O}(m^4)
\label{expansion}
\end{eqnarray}
where we have used (\ref{Landau_entropy}) and the coefficients of the expansion are given by
\begin{equation}
\alpha_k\equiv\frac{1}{k!}\left.\frac{\partial^k}{\partial \varepsilon^k}\alpha_s(\bar{n}_\sub{g},
\varepsilon,\eta)\right|_{\varepsilon=\bar{\varepsilon}},
\qquad \alpha=a,b,c.
\end{equation}
Fixing the temperature to that of the state at which $\varepsilon=\bar{\varepsilon}$ and $m=0$, such that
\begin{equation}
\frac{1}{\tau}=\left.\frac{\partial }{\partial \varepsilon}\hat{s}(m,\varepsilon,\eta)\right|_{m=0,\varepsilon=\bar{\varepsilon}}=a_1,
\label{beta}
\end{equation}
and minimizing (\ref{expansion}) with respect to $q$ yields
\begin{equation}
q=-\frac{b_1}{2 a_2}m-\frac{4a_2^2c_1-4a_2b_1b_2+3a_3b_1^2}{8a_2^3}m^2 +\mathcal{O}(m^3),
\label{q}
\end{equation}
as well as a second solution given by
\begin{equation}
q_2=-\frac{2a_2}{3a_3}-\frac{2b_2}{3a_3}m -q.
\end{equation}
We do not consider the solution $q_2$ because it does not vanish at $m=0$.
Therefore, using equations (\ref{beta}) and (\ref{q}) in (\ref{expansion}) and replacing the latter in (\ref{free_energy2}) gives
\begin{eqnarray}
\fl\hat{\varphi}(m,\tau,\eta)=&\frac{\bar{\varepsilon}}{\tau}-a_0-b_0m+\left(\frac{b_1^2}{4a_2}-c_0\right)m^2\nonumber\\
&+\left(\frac{4a_2^2b_1c_1-2a_2b_1^2b_2+a_3b_1^3}{8a_2^3}-d_0\right)m^3 +\mathcal{O}(m^4). 
\label{free_energy_lowest}
\end{eqnarray}
By writing the free energy as
\begin{eqnarray}
\fl\hat{\varphi}(m,\tau,\eta)=&a_\varphi(\bar{n}_\sub{g},\tau,\eta)+b_\varphi(\bar{n}_\sub{g},\tau,\eta)m
+c_\varphi(\bar{n}_\sub{g},\tau,\eta)m^2
+d_\varphi(\bar{n}_\sub{g},\tau,\eta)m^3\nonumber\\
&+e_\varphi(\bar{n}_\sub{g},\tau,\eta)m^4+\mathcal{O}(m^5),
\label{Landau_free_energy}
\end{eqnarray}
one identifies the coefficients
\begin{eqnarray}
&a_\varphi(\bar{n}_\sub{g},\tau,\eta)=\frac{\bar{\varepsilon}}{\tau}-a_0\label{a_varphi},\\
&b_\varphi(\bar{n}_\sub{g},\tau,\eta)=-b_0,\label{b_varphi}\\
&c_\varphi(\bar{n}_\sub{g},\tau,\eta)=\frac{b_1^2}{4a_2}-c_0\label{c_varphi},\\
&d_\varphi(\bar{n}_\sub{g},\tau,\eta)=\frac{4a_2^2b_1c_1-2a_2b_1^2b_2+a_3b_1^3}{8a_2^3}-d_0\label{d_varphi},
\end{eqnarray}
where the mean energy $\bar{\varepsilon}$ must be taken as a function of $\tau$ whose dependence is obtained through (\ref{beta}).
Since $\hat{s}(m,\bar{\varepsilon},\eta)=a_0+b_0m+c_0m^2+d_0m^3+\mathcal{O}(m^4)$, the previous procedure provides the
firsts terms of the Landau expansion of the canonical free energy as functions of the coefficients of the expansion of the
microcanonical entropy at a certain energy $\varepsilon=\bar{\varepsilon}$.
We observe that the coefficients $c_\varphi$ and $d_\varphi$ do not vanish at the same critical conditions that $c_0$ and
$d_0$ do. Hence, the critical point in the canonical ensemble will be different from the corresponding one in the microcanonical
ensemble.

\begin{figure}
\centering
\includegraphics[scale=1]{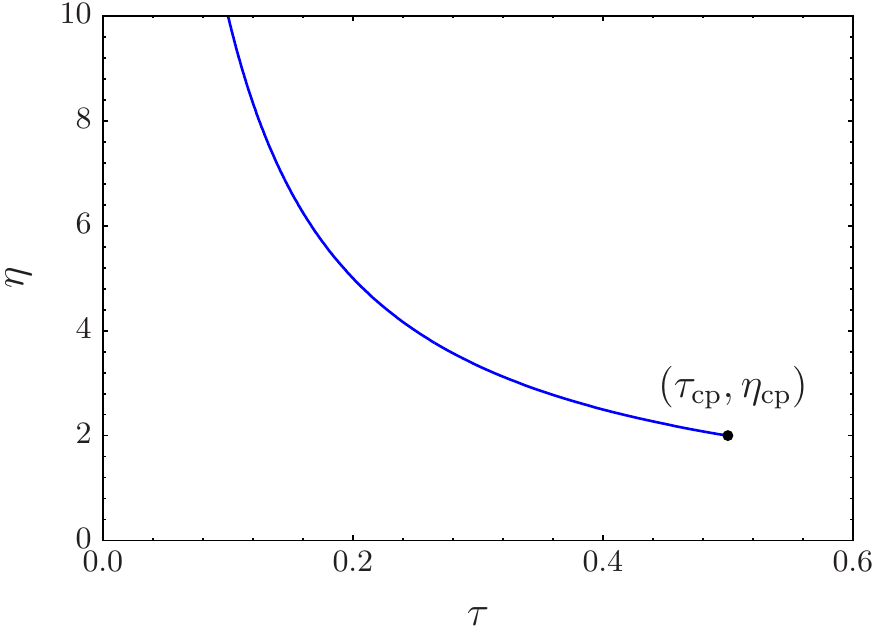} 
\caption{Phase diagram in the $(\tau,\eta)$ plane, showing the transition line in the canonical ensemble. The plot shows the curve
$\eta(\tau)$, which represents the points defined by the
reduced temperature $\tau$ and reduced volume $\eta$ at which first-order phase transitions take place in the canonical ensemble.
The transition line terminates at a critical point represented by the point $(\tau_\sub{cp},\eta_\sub{cp})$. The values of the critical
parameters are $\tau_\sub{cp}=1/2$ and $\eta_\sub{cp}=2$.}
\label{can_phasediagram_beta}
\end{figure}

We highlight that we have started from a generic Landau expansion of the entropy.
For Thirring's model $\bar{\varepsilon}$ is given by (\ref{mean_energy}) and the coefficients in the microcanonical ensemble by
equations (\ref{a_s})-(\ref{e_s}), so that using equations (\ref{b_varphi}), (\ref{c_varphi}), and (\ref{d_varphi}) one obtains
\begin{eqnarray}
\label{coeffcan_b}
&b_\varphi(\bar{n}_\sub{g},\tau,\eta)=\frac{2\left(1-\bar{n}_\sub{g}\right)}{\tau}-\ln\left(\frac{1-\bar{n}_\sub{g}}
{\bar{n}_\sub{g}}\right)-\eta,\label{b_varphi_2}\\
\label{coeffcan_c}
&c_\varphi(\bar{n}_\sub{g},\tau,\eta)=\frac{Q_\tau(\bar{n}_\sub{g})}{\tau\left(1-\bar{n}_\sub{g}\right)\bar{n}_\sub{g}},
\label{c_varphi_2}\\
\label{coeffcan_d}
&d_\varphi(\bar{n}_\sub{g},\tau,\eta)= \frac{1}{6} \left[-\frac{1}{\bar{n}_\sub{g}^2}+\frac{1}{(1-\bar{n}_\sub{g})^2}\right],
\label{d_varphi_2}
\end{eqnarray}
with
\begin{equation}
Q_\tau(\bar{n}_\sub{g})\equiv\bar{n}_\sub{g}^2-\bar{n}_\sub{g}+\frac{\tau}{2}.
\label{polynomial_Q}
\end{equation}
The equilibrium states in the canonical ensemble require the conditions $b_\varphi(\bar{n}_\sub{g},\tau,\eta)=0$,
defining $\bar{n}_\sub{g}=\bar{n}_\sub{g}(\tau,\eta)$, and $c_\varphi(\bar{n}_\sub{g},\tau,\eta) \ge 0$.

\subsection{Canonical critical point}

As it happens in the microcanonical ensemble in the $(\varepsilon,\eta)$ plane, the canonical phase diagram
exhibits in the $(\tau,\eta)$ plane a line of first-order phase transition that ends at a critical point,
here specified by the reduced temperature and
volume $\tau_\sub{cp}$ and $\eta_\sub{cp}$, respectively, for which the fraction of free particles is denoted by $n_\sub{cp}$.
The critical parameters can now be obtained by solving the system of equations
\begin{equation}
\eqalign{b_\varphi(n_\sub{cp},\tau_\sub{cp},\eta_\sub{cp})=0,\\
c_\varphi(n_\sub{cp},\tau_\sub{cp},\eta_\sub{cp})=0,\\
d_\varphi(n_\sub{cp},\tau_\sub{cp},\eta_\sub{cp})=0.
}
\label{cano_system_equations}
\end{equation}
Since $d_\varphi$ depends only on $\bar{n}_\sub{g}$, one immediately obtains that
the critical point can occur only for $n_\sub{cp}=\bar{n}_\sub{g} = 1/2$. Substituting it in equation (\ref{coeffcan_c}) or
equation (\ref{polynomial_Q}) one then finds that the critical temperature is $\tau_\sub{cp}=1/2$. Finally, replacing these
values in equation (\ref{coeffcan_b}) we get that $\eta_\sub{cp}=2$. However, it is useful to consider the discriminant of the
quadratic polynomial $Q_\tau(\bar{n}_\sub{g})$, given in equation (\ref{polynomial_Q}), following a procedure
analogous to that used in the microcanonical case, where the discriminant of the quartic polynomial $P_\varepsilon(\bar{n}_\sub{g})$
was studied. The discriminant of $Q_\tau(\bar{n}_\sub{g})$ takes the form $\Delta_Q(\tau)=1-2\tau$, and we know that the critical
temperature satisfies $\Delta_Q(\tau_\sub{cp})=0$, giving $\tau_\sub{cp}=1/2$. In this case, the roots of $Q_\tau(\bar{n}_\sub{g})$,
given by
\begin{eqnarray}
n_1(\tau)=\frac{1}{2}-\frac{\sqrt{\Delta_Q(\tau)}}{2}, \\
n_2(\tau)=\frac{1}{2}+\frac{\sqrt{\Delta_Q(\tau)}}{2},
\label{roots_can_bis}
\end{eqnarray}
coincide and are equal to $n_\sub{cp}=1/2$. The last expressions also show that for $\tau>\tau_\sub{cp}$, $Q_\tau(\bar{n}_\sub{g})$
has no real roots and, hence, the second order
coefficient $c_\varphi$ does not vanish. This means that in such a case the condition $b_\varphi(\bar{n}_\sub{g},\tau,\eta)=0$
defines only one state of equilibrium, and, therefore, phase transitions can only occur if $\tau<1/2$.

We emphasize that, for Thirring's model, the coefficients (\ref{b_varphi_2}), (\ref{c_varphi_2}), and (\ref{d_varphi_2}) of the
Landau expansion can be alternatively obtained from the free energy (\ref{free_energy}), instead of the method we employed here.
In fact, from (\ref{free_energy}), the remaining coefficient of (\ref{Landau_free_energy}) takes the form
\begin{equation}
e_\varphi(\bar{n}_\sub{g},\tau,\eta)= \frac{1}{12} \left[\frac{1}{\bar{n}_\sub{g}^3}+\frac{1}{(1-\bar{n}_\sub{g})^3}\right].
\label{e_varphi_2}
\end{equation}
However, taking into account that the expansions (\ref{Landau_entropy}) and (\ref{expansion}) do not depend on
the model, some general conclusions can be obtained from this method. From equation (40) one sees that at the canonical critical
point, where $c_\varphi=0$, $c_0$ is different from zero if $b_1\neq0$, implying that the two critical points do not coincide in
general, regardless of the model. Of course, for a particular model, the Landau expansion may present a symmetry with respect to
the order parameter that enforces the condition $b_1=0$~\cite{Cohen_2012}; here we consider that there is no such a symmetry.
Furthermore, it is important to stress that, according to the Landau theory, we are assuming analyticity of the free energy at
the canonical critical point. Analyticity here can be assumed because the system is nonadditive and for these systems actually
there is no phase separation at the transition line. Therefore, this discussion does not apply to short-range interacting systems,
since these systems do undergo phase separation at a first-order transition, which, in addition, occurs under the same conditions
in the different ensembles.

\section{Microcanonical and canonical phase diagrams}
\label{phase_diagrams}

\begin{figure}
\centering
\includegraphics[scale=1]{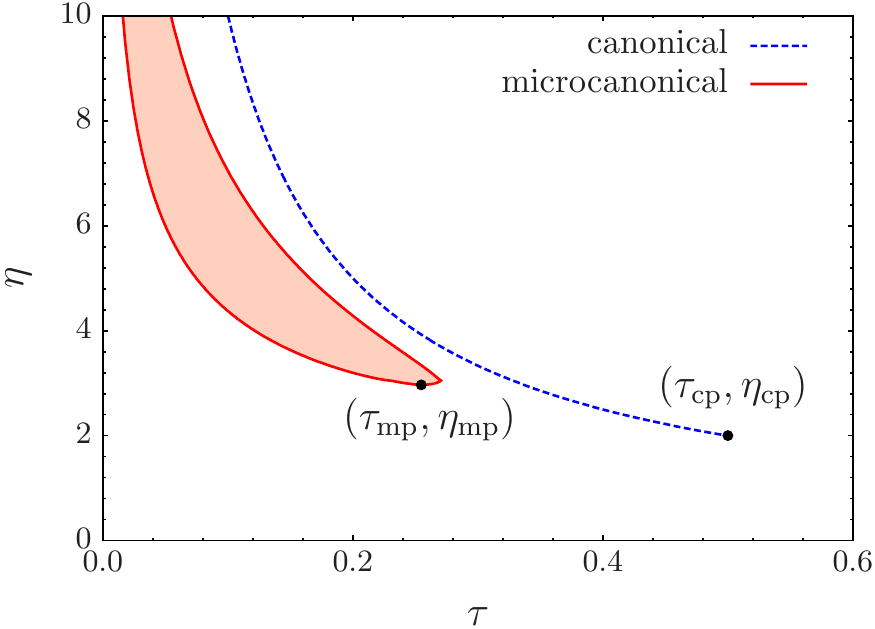} 
\caption{Comparison of the microcanonical and canonical phase diagrams. The reduced volume $\eta$ is shown as a function of the
temperature $\tau$ at the transition line of the two ensembles. In the microcanonical ensemble, there are two temperature branches
that join at the critical point. The temperatures within the region between these two branches are forbidden in the microcanonical
ensemble. In the canonical ensemble, $\tau$ is a control parameter and, thus, has no discontinuity at the transition line.  The
critical parameters are $\tau_\sub{mp}\simeq 0.2547$, $\eta_\sub{mp}\simeq 2.969$, $\tau_\sub{cp}=1/2$ and $\eta_\sub{cp}=2$.} 
\label{micro_can_phasediagram_beta}
\end{figure}

In this section we will draw a comparison between the microcanonical and canonical phase diagrams.
Since the transition is first-order in both ensembles, the two equilibrium configurations associated to the transition are
characterized by a jump in $\bar{n}_\sub{g}$ and, hence, in the associated thermodynamic properties (those which are not
control parameters).

In figure \ref{micro_phasediagram_e}, we plotted the microcanonical transition line $\eta(\varepsilon)$. This line indicates
the points $(\varepsilon,\eta)$ at which phase transition takes place, i.e., when the entropy reaches the same value at the two maxima.
We emphasize that both $\varepsilon$ and $\eta$ are control parameters in the microcanonical ensemble.
In figure \ref{can_phasediagram_beta}, we showed the canonical transition line $\eta(\tau)$ containing the points
$(\tau,\eta)$ at which a phase transition occurs in the canonical ensemble, corresponding to the coincidence of the two free
energy minima. We recall that $\tau$ and $\eta$ are the control parameters  in this ensemble.

On the one hand, in the microcanonical case, the jump in $\bar{n}_\sub{g}$ produces a jump in the temperature. This can be seen
in figure \ref{micro_can_phasediagram_beta} in the microcanonical phase diagram in the $(\tau,\eta)$ plane: we plot $\eta$ at the
transition line as a function of the microcanonical temperature $\tau$. The phase diagram has two branches starting at small
temperatures and high $\eta$ that join smoothly at the critical point. According to (\ref{inverse_temperature_micro}), the
microcanonical temperature at the critical point is $\tau_\sub{mp}\simeq 0.2547$. We highlight that the temperatures between the
two branches of the phase diagram are forbidden for the system in the microcanonical ensemble. In addition, for comparison purposes,
in figure \ref{micro_can_phasediagram_beta} the canonical phase diagram is also shown, where $\tau$ is a control parameter and thus
has no discontinuity. We observe that the microcanonical and canonical critical points are far from each other and
that the temperatures corresponding to the canonical phase transition are allowed in the microcanonical ensemble.

\begin{figure}
\centering
\includegraphics[scale=1]{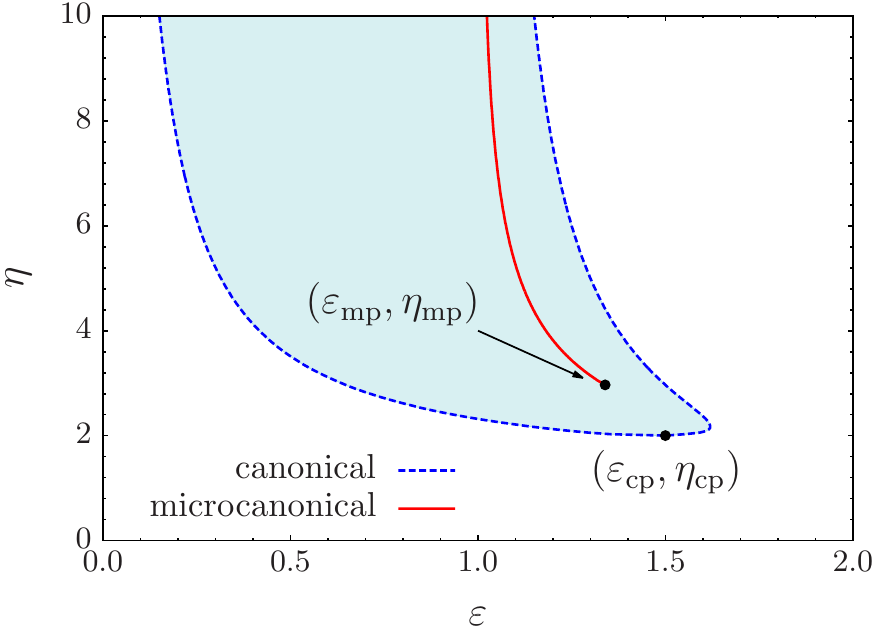} 
\caption{Comparison of the microcanonical and canonical phase diagrams. The reduced volume $\eta$ is shown as a function of the
reduced energy $\varepsilon$ at the transition line of the two ensembles. In the canonical ensemble, there are two energy branches
that join at the critical point. The energies within the region between these two branches are forbidden in the canonical ensemble.
In the microcanonical ensemble, $\varepsilon$ is a control parameter and, thus, has no discontinuity at the transition line.
The critical parameters are $\varepsilon_\sub{mp}\simeq 1.339$, $\eta_\sub{mp}\simeq 2.969$, $\varepsilon_\sub{cp}=3/2$ and
$\eta_\sub{cp}=2$.}
\label{micro_can_phasediagram_e}
\end{figure}

On the other hand, in the canonical ensemble the jump in $\bar{n}_\sub{g}$ at the transition produces a jump in the energy, as
shown in figure~\ref{micro_can_phasediagram_e} in the phase diagram in the $(\varepsilon,\eta)$ plane. In this diagram, there
are two energy branches that join smoothly at the canonical critical point, the energy at this point being $\varepsilon_\sub{cp}=3/2$.
Moreover, due to the jump, the values of the energy between the two branches are forbidden in the canonical ensemble. We finally
observe that the energies at the transition line in the microcanonical ensemble, also shown in figure~\ref{micro_can_phasediagram_e},
lie within the region of forbidden canonical energies.

To get a clearer picture of the behavior of the system when the energy is a control parameter, in comparison
with the situation in which the system is in contact with a thermostat at fixed temperature, in figure \ref{caloric} we show several
caloric curves in both the microcanonical and canonical ensembles. These curves are shown for different values of the reduced
volume $\eta$. When $\eta\leq\eta_\sub{cp}$ the temperature-energy relation $\tau(\varepsilon,\eta)$ is invertible --in the sense
that $\varepsilon(\tau,\eta)$ can be unequivocally obtained from it--, and the microcanonical and canonical ensembles are equivalent.
Notice that in this case the Legendre-Fenchel transform (\ref{Legendre-Fenchel}) reduces to the usual Legendre transform.
For values of $\eta$ such that the temperature-energy relation is not invertible in the microcanonical ensemble, the system undergoes
a first-order phase transition in the canonical ensemble.
Moreover, for $\eta>\eta_\sub{mp}$, the microcanonical phase transition is always jumped over by the transition in the canonical
ensemble. This is, of course, in agreement with the fact that the microcanonical critical point ($\eta=\eta_\sub{mp}$) lies in the
region of forbidden energies in the canonical ensemble.

\begin{figure}
\centering
\includegraphics[scale=1]{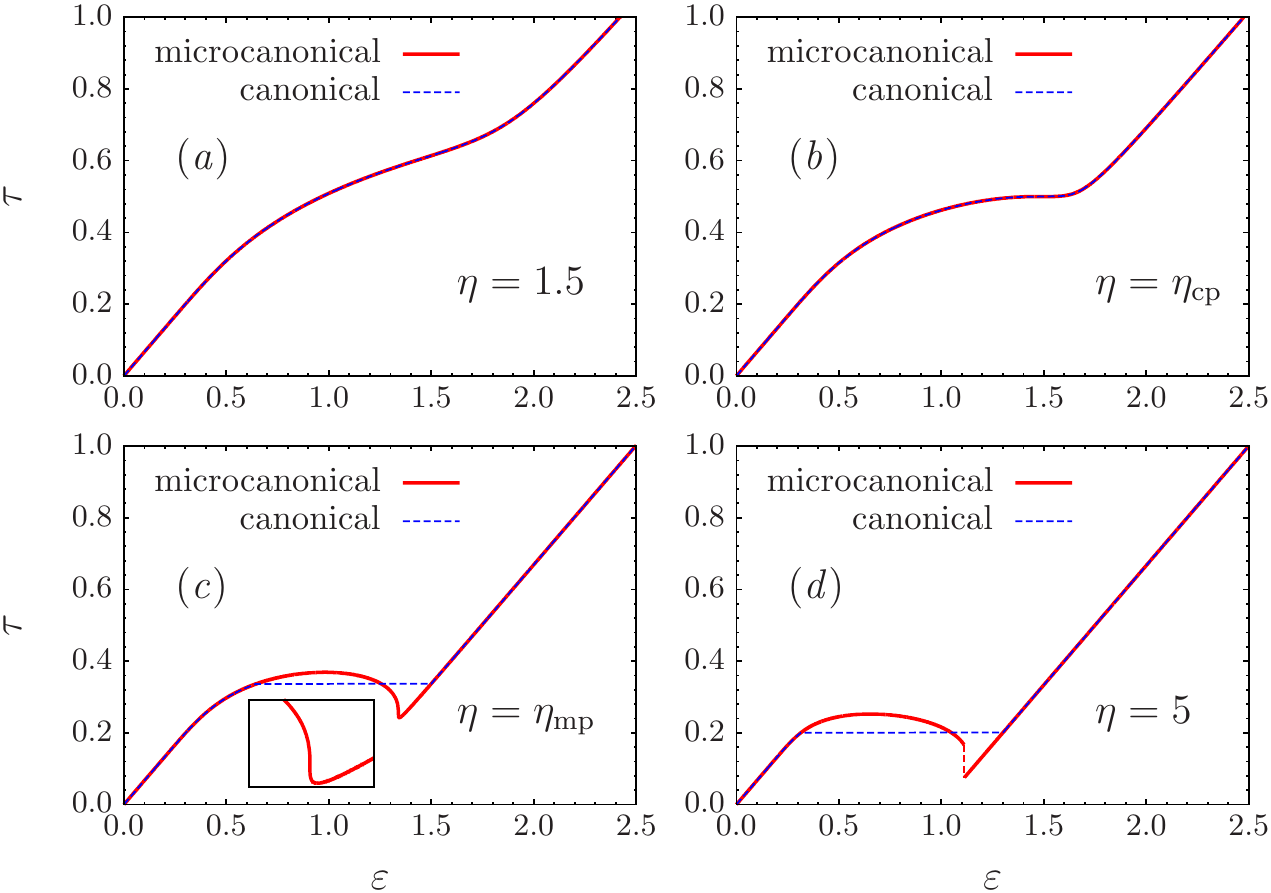} 
\caption{Caloric curves in the microcanonical and canonical ensembles for several values of the reduced volume $\eta$.
For $\eta\leq\eta_\sub{cp}$ the two ensembles are equivalent, as shown in (\textit{a}) and (\textit{b}). In (\textit{c}) and
(\textit{d}), a region of nonconcave entropy, containing a region with negative specific heat, in the microcanonical ensemble appears for $\eta>\eta_\sub{cp}$, which is jumped
over by a first order transition in the canonical ensemble. In (\textit{c}), the curve has a vertical tangent when
approaching from both the left and the right to the critical energy $\varepsilon_\sub{mp}$, occurring just before the local
minimum (see the enlargement in the inset). For $\eta>\eta_\sub{mp}$, in (\textit{d}), the system develops a
temperature jump in the microcanonical ensemble. This jump is denoted with a red dashed line.}
\label{caloric}
\end{figure}

We emphasize that at the canonical critical point the specific heat diverges in both the microcanonical and canonical ensembles. This state of the system is described exactly in the same way in the two ensembles. However, while at $\tau=\tau_\sub{cp}$ and $\eta=\eta_\sub{cp}$ there is a second-order phase transition in the canonical ensemble (that becomes first-order for $\eta>\eta_\sub{cp}$ at the corresponding $\tau$), at this point there is no transition in the microcanonical ensemble.
To see that the mere presence of a diverging specific heat (or a vanishing derivative of the curve $\tau$ vs. $\varepsilon$) is not sufficient to say that there is a phase transition in the microcanonical ensemble, consider for instance the caloric curve for
a value of $\eta$ such that $\eta_\sub{cp}<\eta<\eta_\sub{mp}$, which corresponds to a situation between (\textit{b}) and (\textit{c}) in
figure~\ref{caloric}. 
This curve is continuous with continuous derivative, but with a region of negative specific heat located between the two values of $\varepsilon$ at which the curve $\tau$ vs. $\varepsilon$ has zero derivative. The system does not undergo qualitative changes when passing through any of the points of this curve, which can be achieved by slightly modifying the control parameters in the neighborhood of a given point.
The canonical critical point is, for the microcanonical ensemble, just the first point where a vanishing value of the derivative of $\tau$ as a function of $\varepsilon$ appears.
In other words, since the microcanonical specific heat can be written as
\begin{equation}
c_\sub{micro} = -\left(\frac{\partial s}{\partial \varepsilon}\right)^2\left(\frac{\partial^2 s}{\partial \varepsilon^2}\right)^{-1},
\end{equation}
an inflection point in $s(\varepsilon)$ with vanishing second derivative may produce a diverging microcanonical specific heat. 

It is interesting to note that the regions of ensemble inequivalence and the occurrence of phase transitions in both ensembles
can be deduced from singular points in the $s(\varepsilon,\eta)$ curve or, equivalently, from the microcanonical
temperature-energy relation $\tau(\varepsilon,\eta)$. We identify two different codimension 1 singularities, as classified
in \cite{Bouchet_2005}. Notice that here $\eta$ is the (only one) parameter, in addition to the energy, that can produce a change
in the structure of $s(\varepsilon,\eta)$ or the caloric curve. These singular points can be observed in figure~\ref{caloric},
as we discuss in what follows. At $\eta=\eta_\sub{cp}$, a singularity arises due to convexification, in which a point with
horizontal tangent appears in the curve $\tau(\varepsilon,\eta_\sub{cp})$ (the entropy is concave at this point). This corresponds
to the canonical critical point. The second singularity occurs at $\eta=\eta_\sub{mp}$, which is a maximization singularity, in
that a point with vertical tangent appears in the curve $\tau(\varepsilon,\eta_\sub{mp})$. Such a point correspond to the
microcanonical critical point.

\section{Conclusions}

We have studied the phase diagrams of Thirring's model~\cite{Thirring_1970} in both the microcanonical and canonical ensembles.
Due to the nonadditive character of the system, these two ensembles are not equivalent and the corresponding phase diagrams are
different from each other. Using the Landau theory of phase transitions, as done in~\cite{Cohen_2012}, the coefficients of the
Landau expansion of the canonical free energy can be written in terms of the coefficients of the microcanonical entropy, which
permits an analysis of the critical conditions in which these coefficient vanish. Hence, the critical point at which each
first-order transition line terminates can be computed exactly, evincing that they are indeed different. 
Since the analysis was performed from generic expansions for the entropy and the free energy, it can be inferred that, in general,
the critical points in the two ensembles are different.
As a difference with respect to~\cite{Cohen_2012}, here we have considered that there are no
symmetries restricting the coefficients of the Landau expansion of the corresponding thermodynamic potentials.
Moreover, the comparison of the two phase diagrams shows that the energies at which phase transitions take place in the
microcanonical ensemble are not allowed in the canonical ensemble. Conversely, the temperatures at which the transitions take place
in the canonical ensemble are accessible to the microcanonical equilibrium configurations.  

Furthermore, in this case, the microcanonical entropy and the canonical free energy can be written as a maximization and minimization
problem, respectively, with respect to a variable $n_\sub{g}$ (here representing the fraction of free particles), and the
expansion parameter in the Landau expressions can be written as
$m=n_\sub{g}-\bar{n}_\sub{g}$, where $\bar{n}_\sub{g}$ is a reference value. In addition, the Landau expansions are
complete, at least to the fourth order, and the second order coefficient in the two ensembles is a rational function of the
form $f_\lambda(\bar{n}_\sub{g})/g_\lambda(\bar{n}_\sub{g})$, where $f_\lambda(\bar{n}_\sub{g})$ and $g_\lambda(\bar{n}_\sub{g})$
are polynomials of the variable $\bar{n}_\sub{g}$ that depend on the parameter $\lambda$. The parameter $\lambda$ is the energy in
the microcanonical case and the temperature in the canonical one. Thus, we have shown that, in general, when this conditions are
met, the critical point in each ensemble can be obtained by studying the zeros of the discriminants $\Delta_f(\lambda)$ associated
to the polynomials $f_\lambda(\bar{n}_\sub{g})$.

\ack

We thank O Cohen and D Mukamel for fruitful discussions.
I L acknowledges financial support through an FPI scholarship (Grant No. BES-2012-054782) from
the Spanish Government. This work was partially supported by the Spanish Government under Grant No. FIS2011-22603.
We also thank the Galileo Galilei Institute for Theoretical Physics for the hospitality and the INFN for partial support
during the completion of this work.

\end{document}